\documentclass[reprint,amsmath,amssymb,aip,apl]{revtex4-1edit}
\usepackage{graphicx}
\usepackage{dcolumn}

\begin{document}

\title{All-silicon sub-Gb/s telecom detector with low dark current and high quantum efficiency on chip}
\author{Takasumi Tanabe,\textsuperscript{1, 2}}
\email{takasumi@nttbrl.jp}
\affiliation{NTT Basic Research Laboratories, NTT Corporation, 3-1 Morinosato Wakamiya, Atsugi, Kanagawa 243-0198, Japan}
\affiliation{CREST-JST, 4-1-8 Honmachi, Kawaguchi, Saitama 332-0012, Japan}
\author{Hisashi Sumikura,\textsuperscript{1}}
\affiliation{NTT Basic Research Laboratories, NTT Corporation, 3-1 Morinosato Wakamiya, Atsugi, Kanagawa 243-0198, Japan}
\author{Hideaki Taniyama,\textsuperscript{1, 2}}
\affiliation{NTT Basic Research Laboratories, NTT Corporation, 3-1 Morinosato Wakamiya, Atsugi, Kanagawa 243-0198, Japan}
\affiliation{CREST-JST, 4-1-8 Honmachi, Kawaguchi, Saitama 332-0012, Japan}
\author{Akihiko Shinya,\textsuperscript{1, 2}}
\affiliation{NTT Basic Research Laboratories, NTT Corporation, 3-1 Morinosato Wakamiya, Atsugi, Kanagawa 243-0198, Japan}
\affiliation{CREST-JST, 4-1-8 Honmachi, Kawaguchi, Saitama 332-0012, Japan}
\author{Masaya Notomi \textsuperscript{1, 2}}
\email{notomi@nttbrl.jp}
\affiliation{NTT Basic Research Laboratories, NTT Corporation, 3-1 Morinosato Wakamiya, Atsugi, Kanagawa 243-0198, Japan}
\affiliation{CREST-JST, 4-1-8 Honmachi, Kawaguchi, Saitama 332-0012, Japan}

\date{\today}

\begin{abstract}
 We demonstrate channel selective 0.1-Gb/s photo-receiver operation at telecom wavelength using a silicon high-$Q$ photonic crystal nanocavity with a laterally integrated \textit{p-i-n} diode.  Due to the good crystal property of silicon the measured dark current is only 15~pA.  The linear and nonlinear characteristics are investigated in detail, in which we found that the photo-current is enhanced of more than $10^5$ due to the ultrahigh-$Q$ ($Q\simeq 10^5$).  With the help of two-photon absorption, which is visible at a surprisingly low input power of $10^{-8}$~W, the quantum efficiency of this device reaches $\sim 10$\%.
\end{abstract}

\maketitle
Silicon (Si) has long been the main platform in electronics, but it is also an excellent material for use in photonics.  Its high index allows us to fabricate high-$Q$ nanocavities\cite{Tanabe2007np, Noda2007} and low-loss waveguides.\cite{Notomi2007}  However, Si is not an efficient material in terms of telecom-light detection.  Various approaches have been attempted to overcome this problem including the integration of germanium on Si.\cite{Yin2007, Kang2009}  Nevertheless, reducing the dark current remains a challenge because of the lattice mismatch between the materials.  All-Si approaches have also been attempted by introducing ion-implantation,\cite{Liu2006, Geis2009} but the same problem remains owing to the presence of defects.  Pure Si detectors should make it possible to achieve low-noise and ease of fabrication, but we need to employ two-photon absorption (TPA), which has a low coefficient.  TPA detection has been demonstrated using a \textit{p-i-n} integrated Si waveguide,\cite{Liang2002} but it required a high optical input.  The key to compensating for the low TPA coefficient is to strengthen the interaction between light and matter by tightly confining the light.  A high-$Q$ photonic crystal (PhC) nanocavity is a good candidate because it allows us to achieve a high photon density even with a very low input power.  Indeed, a numerical study has shown that a \textit{p-i-n} Si PhC nanocavity is a good candidate for a high-speed high-efficient photo receiver.\cite{Abad2009}

\begin{figure}[b]
 \begin{center}
  \includegraphics[width=3in]{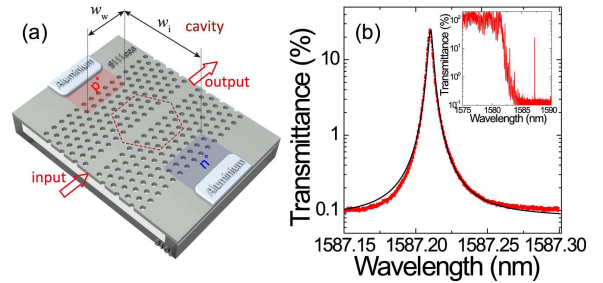}
  \caption{(Color online) (a) Schematic illustration of a \textit{p-i-n} PhC nanocavity.  The lattice constant $a$, hole radius $r$, slab thickness $t$, $w_\mathrm{i}$, and $w_\mathrm{w}$ are 420~nm, 108~nm, 204~nm, 8.72~$\mathrm{\mu m}$, and 8.4~$\mathrm{\mu m}$, respectively.  The input and output waveguides with widths of $1.05a$ are in-line connected with the cavity through barrier line defects with widths of $0.98a$.  The length of the barrier line defect is $d=11a$.  (b) Optical transmittance spectrum.  The inset shows the overview spectrum.}
  \label{fig1}
 \end{center}
\end{figure}

\begin{figure}[tb]
 \begin{center}
  \includegraphics[width=3.5in]{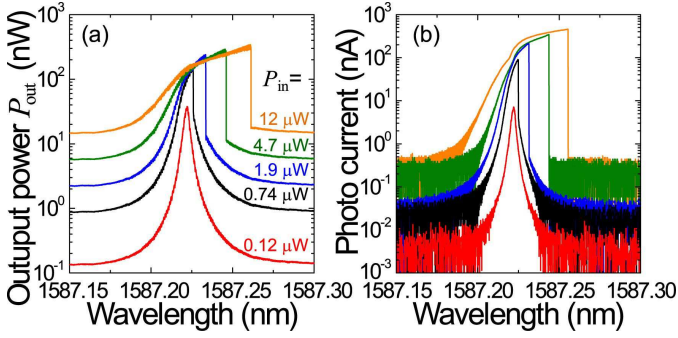}
  \caption{(Color online) (a) Optical spectra at different input powers at reverse $-3$~V.  $P_\mathrm{in}$ and $P_\mathrm{out}$ are the powers at the PhC waveguides.  (b) Current spectra corresponding to (a).}
  \label{fig2}
 \end{center}
 \end{figure}

We devised a high-$Q$ width-modulated line defect PhC nanocavity with an integrated lateral \textit{p-i-n} structure \cite{Tanabe2009oe} to improve the efficiency of the TPA and one-photon absorption (OPA) carrier generation.  Figure~\ref{fig1}(a) shows a schematic image of the device.  The fabrication process is described elsewhere.\cite{Tanabe2009oe}  Figure~\ref{fig1}(b) shows the optical transmittance spectrum, which gives a $Q$ of $4.3\times10^5$ and a transmittance $T_0$ of 24\%.  This corresponds to an intrinsic $Q$ of $8.4\times10^5$.
We apply a reverse $-3$~V to the \textit{p-i-n} diode and inject CW laser light whose wavelength is swept from short to long.  Figure~\ref{fig2}(a) shows the transmitted optical spectra for different input laser powers $P_\mathrm{in}$, where we observe nonlinear modulation resulting from the thermo-optic bistability.\cite{Notomi2005}  At the same time as the optical measurement, we recorded the diode current to obtain the current spectrum, as shown in Fig.~\ref{fig2}(b).  The shape is similar to that in Fig.~\ref{fig2}(a) and a photo-excited current is clearly visible when the input light resonates with the cavity even at the lowest input power.

\begin{figure}[tb]
 \begin{center}
  \includegraphics[width=3.48in]{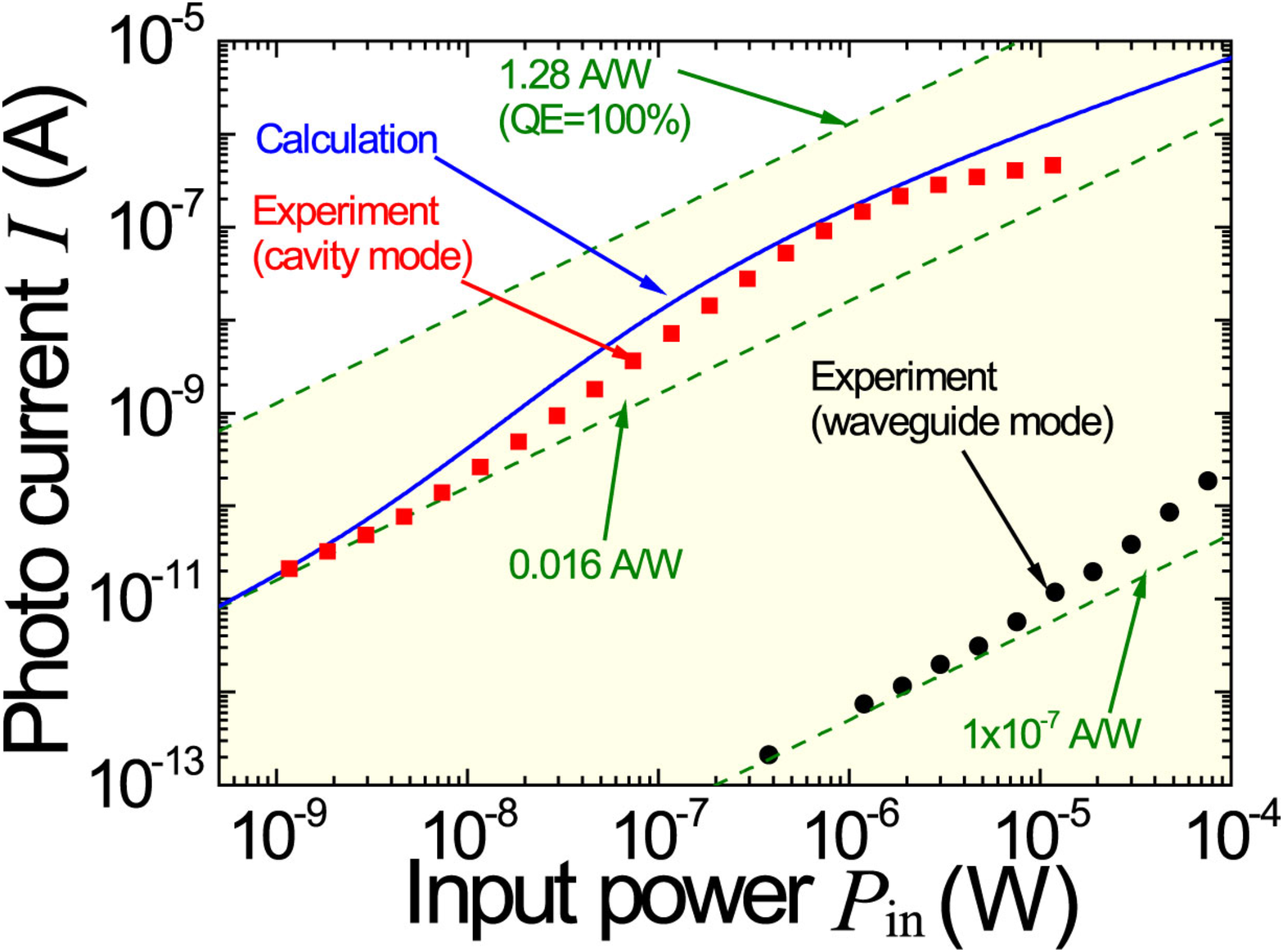}
  \caption{(Color online) Photocurrent ($I$) versus input power at PhC waveguide ($P_\mathrm{in}$).  The red squares are when the input laser light is at the resonance of the cavity.  The black circles are when the input laser wavelength is 1577~nm.  The solid blue line is the calculation.  The dotted green lines show the responsivities at different A/W values.  1.28~A/W corresponds to QE=100\% (one photon generates one electron).}
  \label{fig3}
 \end{center}
\end{figure}

To understand the relationship between the injected light and the current, we plot the peak current $I$ of Fig.~\ref{fig2}(b) as a function of $P_\mathrm{in}$ in Fig.~\ref{fig3}.  We also show the current when we injected 1577-nm CW light.  The light at 1577~nm propagates through the PhC waveguide without resonating with the cavity, because it is above the mode-gap barrier frequency as shown in the inset of Fig.~\ref{fig1}(b).  From the plots in Fig.~\ref{fig3}, we immediately know that the formation of a high-$Q$ mode significantly improves the generation of photocurrent in a Si \textit{p-i-n} diode by a factor of $> 10^5$, which enables us possible to detect very weak CW light.

Next, we investigate the linear and nonlinear physics of the device by analysing the square dots of Fig.~\ref{fig3}.  We find that first $I$ increases linearly (Region I:  $P_{\mathrm{in}} < 10^{-8}~\mathrm{W}$), then there is a sub-quadratic increase (Region II: $10^{-8}~\mathrm{W} < P_{\mathrm{in}} < 10^{-6}~\mathrm{W}$), and finally it saturates (Region III:  $10^{-6}~\mathrm{W} < P_\mathrm{in}$).  To understand this in detail, we introduce two rate equations that describe the light energy $u$ stored in the cavity and the generated carrier density $N$.  The equations are,
\begin{eqnarray}
 \frac{du}{dt}&=&\sqrt{T}P_{\mathrm{in}}-\frac{u}{\tau_{\mathrm{ph}}} - \sigma\frac{c}{n}Nu -\frac{2c^2 \beta}{n^2 V_{\mathrm{m}}} u^2  \label{eq1} \\
 \frac{dN}{dt}&=&\frac{2c\beta \lambda}{hn^2 V_{\mathrm{m}}^2}u^2 + \frac{\alpha_{\mathrm{OPA}}\lambda}{hn V_{\mathrm{m}}} u - \frac{N}{\tau_{\mathrm{c}}} - \frac{I}{e V_{\mathrm{m}}}\label{eq2}
\end{eqnarray}
where $\alpha_{\mathrm{OPA}}$ is the coefficient of the OPA, which is not negligible in a weak excitation regime.\cite{Matsuda2009}  $c$, $h$, and $e$ are the light velocity, Planck constant, and electrical charge, respectively.  $n=3.45$ and $\mu=1.45\times10^{-1}~\mathrm{m^2 V^{-1} s^{-1}}$ are the refractive index and electron mobility of Si.  $\phi=3$~V, $\lambda=1587$~nm, $\tau_\mathrm{ph}=0.37$~ns, $d=8.72~\mathrm{\mu m}$, and $S=1.68~\mathrm{\mu m^2}$, are reverse bias voltage, light wavelength, photon lifetime, insulator length, and cross section of insulator, respectively.  These values are obtained by experiment.  $\tau_\mathrm{c}=1$~ns, $\beta=0.9$~cm/GW, and $\sigma=1.45\times10^{-21}~\mathrm{m^2}$ are the carrier lifetime, TPA coefficient, and FCA absorption coefficient, respectively, and the values are taken from references.\cite{Tanabe2007, Rieger2004, Soref1987}  $V_\mathrm{m}=1.73\times10^{-1}~\mathrm{\mu m^3}$ is the mode volume and is obtained from FDTD calculation.  According to the coupled mode theory, $T$ changes when the internal loss of the cavity changes as given by,
\begin{equation}
 T^{1/2} = \frac{T_0^{1/2}}{\tau_{\mathrm{ph}}}\left( \tau_{\mathrm{ph}}^{-1} +\sigma\frac{c}{n}N +\frac{2c^2\beta}{n^2 V_\mathrm{m}} u  \right)^{-1} \label{eq3}
\end{equation}
Wavelength detuning is already taken into account, because the peak current $I$ is obtained when the laser wavelength and the cavity resonance match during wavelength scanning.  The Einstein-Smoluchowski relation gives the $I$ and $N$ values in a steady state as,
\begin{equation}
I=\frac{eS\phi\mu}{d} N \label{eq4}
\end{equation}
Substituting Eqs.~(\ref{eq3}) and (\ref{eq4}) into Eqs.~(\ref{eq1}) and (\ref{eq2}), we can derive an equation that describes the relationship between $P_\mathrm{in}$ and $I$ in the steady state ($du/dt=0$ and $dN/dt=0$). We obtain the graph shown by the solid line in Fig.~\ref{fig3}.  The only fitting parameter is $\alpha_{\mathrm{OPA}}$, and it is given as 0.032~dB/cm.

When we neglect $\alpha_{\mathrm{OPA}}$, the linear increase in Region I disappears.  TPA carrier generation (1st term on the RHS in Eq.~(\ref{eq2})) is needed to make the nonlinear increase in Region II visible.  But without the TPA optical loss (4th term on the RHS in Eq.~(\ref{eq1})), Region II exhibits a quadratic increase and this does not explain the sub-quadratic increase.  These facts indicate that TPA outperforms OPA in Region II.  The saturation of Region III appears when we introduce FCA optical loss (3rd term on the RHS in Eq.~(\ref{eq1})) and this curve provides a better match when we consider Eq.~(\ref{eq3}) instead of using constant $T$.  From these numerical observations, we can conclude  that OPA is visible in Region I, whereas TPA is dominant in Region II.  And FCA appears in Region III.  In other words, nonlinear carrier generation is visible at a $10^{-8}$~W power level in Si, and is made possible by the strong confinement of light.

Here we consider the external quantum efficiency (QE) of our device.  The QE is usually very low in Si at telecom wavelengths because light is transparent.  In fact, the responsivity of the PhC waveguide mode (circles) is only $1\times 10^{-7}$~A/W.  However, the value is increased $10^5$ times (0.016~A/W) by the formation of the high-$Q$ cavity mode in Region I.  In addition, the presence of TPA (Region II) further boosts the QE.  We obtained a QE of 9.7\% at $P_{\mathrm{in}}=1.17$~W.  Considering the wavelength and the material that we used, this value is surprisingly high.  While previous TPA detectors could only detect short high-intensity pulses,\cite{Liang2002} our device even made low-power detection possible.  Furthermore, when we consider the coupling between the waveguide and the cavity [Eq.~(\ref{eq3})], we find that a very large amount ($\sim 44$\%) of the cavity coupled light is absorbed when $P_{\mathrm{in}}=1.17~\mathrm{\mu W}$.

Another important aspect of our Si detector is the small dark current, which enables us to achieve signal detection at an extremely low power.  The maximum measured dark current at room temperature was 15 pA at a reverse bias of $-3$~V.  This value is much smaller than that of other detectors fabricated on Si \cite{Yin2007, Kang2009, Liu2006, Geis2009} and even smaller than that of InGaAs detectors.  Such a small value is achieved because of the good crystal quality of Si and the small dimensions of our \textit{p-i-n} structure.

\begin{figure}[tb]
 \begin{center}
  \includegraphics[width=3.5in]{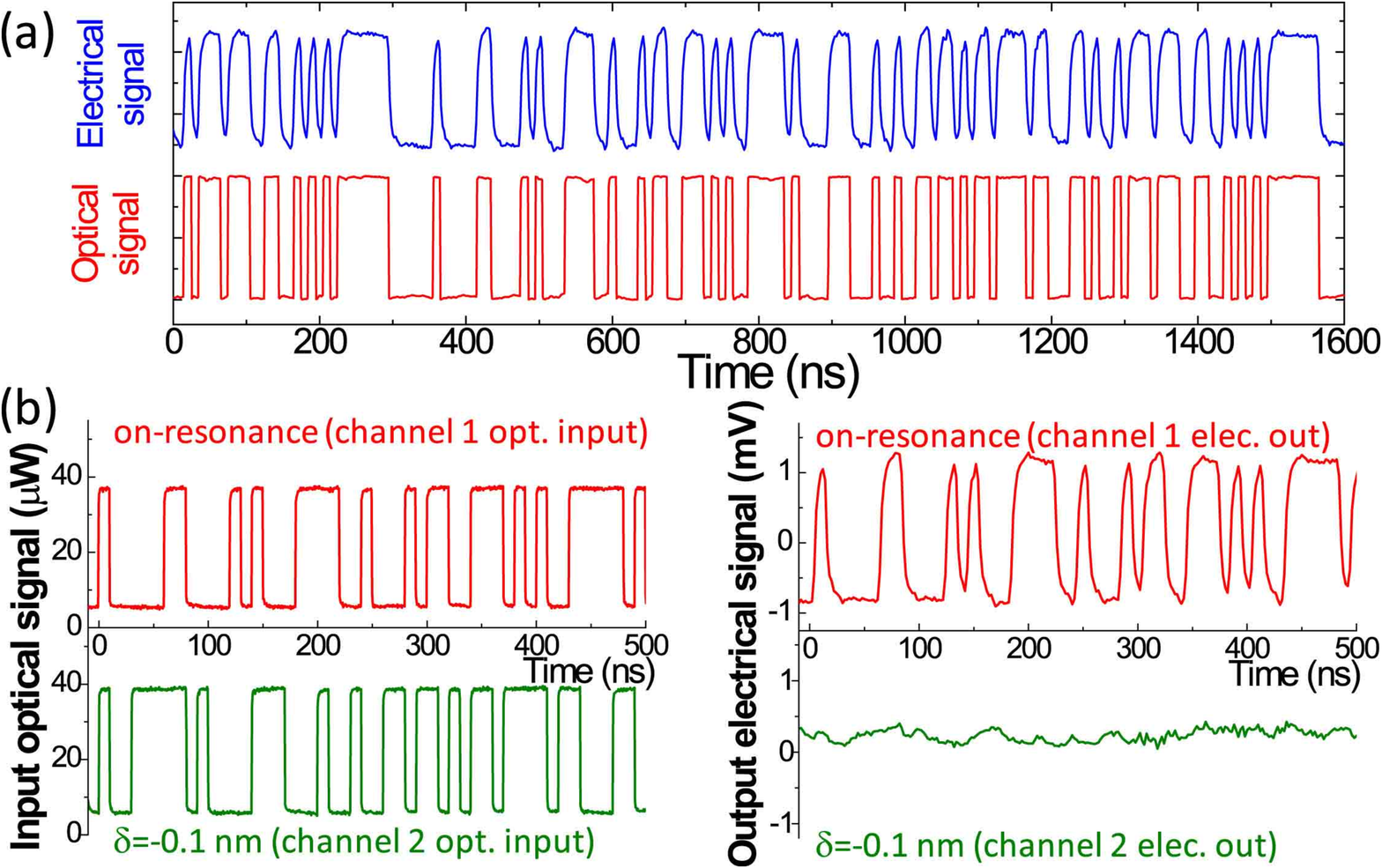}
  \caption{(Color online) (a) 0.1-Gb/s photo-receiver demonstration. The input light has an amplitude of 32~$\mathrm{\mu W}$.  It is a pseudo-random bit sequence with a pattern length of $2^7 -1$.  The amplitude of the electrical output is $\sim 2$~mV.  (b) Input optical signal and the recorded electrical signal for two different input wavelengths.}
  \label{fig4}
 \end{center}
\end{figure}
Finally, we discuss the 0.1~Gb/s photo-receiver operation.  We employed a \textit{p-i-n} PhC nanocavity with the design shown in Fig.~\ref{fig1}(a) except that $d=7a$, and AC-coupled it to an amplifier that has a  transimpedance gain of $5\times10^3$~V/A.  The $Q$, $T_0$ and $\lambda$ were $2.1\times10^4$, $>60$\%, and 1592.4~nm, respectively.  We chose a cavity with a lower $Q$ because it can operate without exhibiting the thermo-optic effect at the higher optical input we used owing to the low transimpedance gain.  When the injected signal wavelength was at the resonance of the cavity, they were detected as shown in Fig.~\ref{fig4}(a).  Figure~\ref{fig4}(b) shows a demonstration at 0.1~Gb/s, with different input laser wavelengths.  Channel 1 is at the resonance and channel 2 is detuned by $-0.1$~nm from the cavity resonance.  Only channel 1 is detected, which constitutes a demonstration of high-speed channel selective photo-detection on a Si chip.  The input optical signal amplitude was $32~\mathrm{\mu W}$, which corresponds to 0.3~pJ/bit at the demonstrated speed.  We may be able to reduce this value, because our Si device has a very small dark current.  In fact, Fig.~\ref{fig3} suggests that the device can detect 12-fJ/bit pulses with a signal rate of 0.1 Gb/s at an optimised $P_\mathrm{in}$ ($1.17~\mathrm{\mu W}$).

Another feature is the small capacitance.  This was achieved because the \textit{p-i-n} structure is small.  The small demensions were possible because we used an ultrasmall PhC cavity.  From the structural dimensions, the estimated capacitance is $9.5\times 10^{-18}$~F.  The small capacitance allows us to obtain a high swing voltage, which may allow us to feed the signal directly into CMOS circuits without using a noisy amplifier.\cite{Miller2009}  To obtain high voltage (and high speed), it is important to have small resistances for the $p$ and $n$ regions.

In terms of operating speed, we recently demonstrated GHz modulation using these devices,\cite{Tanabe2009oe} and an even higher speed has been demonstrated using similar devices.\cite{Xu2007}  Therefore, we believe a GHz or greater speed to be possible by optimizing the electrical design, which may further reduce the J/bit value.  Note that the optical property of the cavity allows the device in Fig.~\ref{fig1}(a) to operate at a speed of 2.7~GHz (given by $\tau_{\mathrm{ph}}^{-1}$) without sacrificing the sensitivity.

In summary, we fabricated a \textit{p-i-n} integrated PhC nanocavity on a Si chip and detected telecom light at a very small dark current and high QE.  We would like to note that this device also paves the way for the development of integrated quantum devices by combining it with single electron transistors.\cite{Fujiwara2008}

We sincerely thank Dr. K. Nishiguchi, Dr. T. Tamamura and Dr. E. Kuramochi for helping with fabrication and for valuable discussions.  We also thanks Dr. Y. Muramoto, Dr. K. Nozaki, Dr. N. Matsuda, Dr. K. Yamada and Dr. A. Fujiwara for fruitful discussions.

\end{document}